\documentclass[sigconf]{acmart}

\usepackage{hyperref}
\usepackage{subcaption}
\usepackage{enumitem}
\usepackage{mathptmx}
\usepackage[11pt]{moresize}
\usepackage{filecontents}

\usepackage{array}
\newcommand*{\origrightarrow}{}

\renewcommand*{\textrightarrow}{\fontfamily{cmr}\selectfont\origrightarrow}

\newcolumntype{M}[1]{>{\centering\arraybackslash}m{#1}}
\newcolumntype{N}{@{}m{0pt}@{}}

\setcopyright{rightsretained}
\renewcommand\footnotetextcopyrightpermission[1]{} 
\begin{document}
\title{Detecting Parkinson's Disease from interactions with a search engine: Is expert knowledge sufficient?}

\author{Liron Allerhand, Brit Youngmann, Elad Yom-Tov}
\affiliation{\institution{Microsoft Research\\ Herzliya, Israel}}
\email{{lironall,t-bryoun,eladyt}@microsoft.com}

\author{David Arkadir}
\affiliation{\institution{Department of Physiology
Hadassah Medical School \\ Jerusalem, Israel}}
\email{Arkadir@hadassah.org.il}

\begin{abstract}
Parkinson's disease (PD) is a slowly progressing neurodegenerative disease with early manifestation of motor signs. Recently, there has been a growing interest in developing automatic tools that can assess motor function in PD patients. Here we show that mouse tracking data collected during people's interaction with a search engine can be used to distinguish PD patients from similar, non-diseased users and present a methodology developed for the diagnosis of PD from these data. 

A main challenge we address is the extraction of informative features from raw mouse tracking data. We do so in two complementary ways: First, we manually construct expert-recommended informative features, aiming to identify abnormalities in motor behaviors. Second, we use an unsupervised representation learning technique to map these raw data to high-level features. Using all the extracted features, a Random Forest classifier is then used to distinguish PD patients from controls, achieving an AUC of $0.92$, while results using only expert-generated or auto-generated features are $0.87$ and $0.83$, respectively. Our results indicate that mouse tracking data can help in detecting users at early stages of the disease, and that both expert-generated features and unsupervised techniques for feature generation are required to achieve the best possible performance. 
\end{abstract}

\maketitle

\section{Introduction}

Parkinson's disease (PD) is the second most prevalent neurodegenerative disorder, affecting $0.3\%$ of the general population and approximatively $1\%$ of people over the age of $60$ \cite{de2006epidemiology}. The physical symptoms of the disease generally appear slowly over time, and often include tremors, rigidity and slowness of movement \cite{de2006epidemiology}.
PD diagnosis relies mostly on clinical judgment of
neurologists evaluating the severity of motor and
non-motor manifestations of the disease \cite{gelb1999diagnostic}.
Recently, there has been a growing interest in developing
automatic tools that can assess motor function in PD patients or in people with high risk of developing PD \cite{giancardo2016computer}. Such tools
can complement physician diagnosis, introducing the potential
for greater screening opportunities and for early diagnosis of the disease.

Technology companies are nowadays capable of assessing the medical condition of their online users, based on users' daily interactions with the company's services (e.g., search engine queries, social media posts). Such interactions include a variety of input signals, such as voice, keystrokes and mouse movements, which are recorded to improve performance and user experience \cite{huang2011no}. These interactions can also be leveraged to reveal the underlying medical conditions of users, based on statistics collected from millions of customers. Past work has indicated the possibility of detecting certain forms of cancer from the text of queries made by people \cite{yom2018clinically,white2017}, or of PD via users interaction \cite{giancardo2016computer,white2018detecting} (see Related Work).

In this work, we present a methodology developed for the diagnosis of PD from the interaction with a search engine results page (SERP). 
Mouse (or cursor) tracking is the use of a software to collect the positions of the users' mouse cursors on the computer or browser page. These data are gathered to obtain richer information on the interaction between the user and a computer or a website, typically to improve the design of an interface \cite{McCay-Peet:2012:SAF:2207676.2207751}, to measure relevance \cite{huang2011no} or, more recently, to estimate search satisfaction, attention and interest \cite{jiang2014searching,arapakis2014user,Lagun:2015:ISA:2766462.2767745}. Here we show that these data make it possible to detect PD, indicating the possibility of using this methodology for identification of users at early stages of PD. 

A main challenge addressed in this work is the extraction of informative features from the raw time-stamped mouse tracking data. We do so in two complementary ways: First, we manually construct informative features suggested by experts, aiming to identify abnormalities in motor behaviors. These include rigidness and shakiness, both primary symptoms of PD. Second, we use an unsupervised representation learning technique. Our technique is based on an approach suggested in \cite{radford2017learning}, which demonstrated a method for learning representation of text
documents, obtaining features corresponding to
high-level concepts. 
In essence, the authors of \cite{radford2017learning} used character-lever predictions to extract high-level features (e.g. sentiment) from documents. Similarly, we predict event-level mouse movements, in order to extract features from mouse movements, and, similarly to \cite{radford2017learning}, we find that a single neuron in our LSTM model predicts whether the person performing the session has PD. 

Our feature extraction problem is, however, inherently different than the one presented in \cite{radford2017learning}. First, time is not a factor when analyzing text. However, mouse events are sampled in a non-uniform manner, and hence the time gap between consecutive mouse events should be taken into account. Second, sampling mouse events is noisy, which results in missing and corrupted events. To overcome the first problem, we predict not only the next mouse position, but also the time stamp of the next event, allowing the algorithm to derive higher level derivatives such as velocity and acceleration as intermediate quantities. To cope with the problem of noisy data, we use robust scaling (see Section \ref{sec:mouse_tracking}). 
One of the contributions of the this work is therefore a generalization of the method proposed in \cite{radford2017learning}, to handle non-uniform sampling and noisy data.

In our experimental study we show the ability to detect PD from mouse tracking. Furthermore, we analyze and compare between the two approaches of feature extraction, discussing the affect of different parameters on performance. 
To the extent of our knowledge, this is the first work that uses representation learning techniques to extract features from mouse tracking data with the goal of identifying the underlying medical state of users (specifically, PD), and believe that our suggested methodology is widely applicable to a range of medical conditions of users.

\vspace{-4px}

\section{Related Work}
Understanding how users interact with the SERP is a fundamental question in information retrieval, bearing on relevance evaluation, search quality, and interface design \cite{huang2011no,Lagun:2015:ISA:2766462.2767745}. Previous work has suggested the use of cursor movements to understand user behavior \cite{arapakis2016predicting}. These data have been successfully used to measure user attention in web search \cite{Huang:2012:USU:2207676.2208591}. Specifically, mouse tracking data was used to infer content salience \cite{Lagun:2015:ISA:2766462.2767745}, improve ranking by estimating the relevance of results \cite{huang2011no}, and dynamically estimate the result that searchers will request next \cite{Diaz:2016:SRP:2911451.2911516}.
These works, while resembling ours in attempting to use mouse tracking signals to understand user behavior, differ in that in our work we attempt to estimate their medical state. 

Another attempt of identifying PD from users' interaction with computers was proposed in \cite{giancardo2016computer}. However, as opposed to our work, the participants were recruited for this task and hence were aware of its purpose. In the current work, the users spontaneously interact with the search engine in different times of the day, providing more natural interactions. 
Moreover, in their work, they only used keyboard data, while here, mouse tracking data is considered as well. 
The authors of \cite{white2018detecting} have recently suggested that people with PD could be distinguished from other people using their mouse movement. In contrast with the current work, the examined control group is consist of the entire population whereas here we consider only spouses of PD patients, and, importantly, only handcrafted features are considered in their models.  

Recently, several works had focused on identifying different types of cancer using users' search queries \cite{yom2018clinically,white2017}. The underlying assumption in these studies, which distinguishes our work, is that some types of cancers manifest themselves in externally recognizable symptoms, which are either unfamiliar to people or relatively benign and so do not cause sufficient alarm. We note that this is not the case with PD, which is recognized by being accompanied by motor symptoms such as rigidness and shakiness. Furthermore, besides user queries, in the current research mouse tracking is leveraged to estimate the underlying medical state as well.

Two common methods for unsupervised features extraction from sequences are reconstructing the input data (e.g., PCA, auto-encoders \cite{guyon2008feature}) and predicting the next step of a given sequence \cite{radford2017learning}.
As mentioned, the latter approach has been recently used to extract features for a sentiment analysis task \cite{radford2017learning}. Interestingly, while the authors of \cite{radford2017learning} found that using a character-level LSTM model to extracts features is highly useful, they also discovered a single neuron performing sentiment detection (i.e., the label) as one of the learned features. As we show, this approach can be generalized to handle non-uniform sampling and noisy data, while the desired label is still learned as one of the features.

\section{Experimental Methodology}
\label{Sec:Exp}

We next explain how mouse tracking data was extracted and used, then shortly describe how features were constructed from these data. Finally, we present the model used in our experimental study.  

\subsection{Data}
We extracted all interactions (mouse movements and keyboard strokes) by anonymized users in the US to the Bing search engine who disclosed having PD by querying for ``I have Parkinson'', ``I have been diagnosed with Parkinson'' or ``my Parkinson...'', between January $2016$ and April $2017$. As Microsoft stores user data only up to $18$ months, we are limited to this period of time. A total of $281$ users made this disclosure. Since PD is a more common disease among the elderly population, as a control group, we used data of $163$ users who queried for ``My husband/wife has Parkinson'', or ``My husband's/wife's Parkinson". Note that this choice of a baseline group also eliminates some demographic and economic differences between users (though usually not gender differences). 
We collected over $800$K user queries and their corresponding data collected (a median of $972$ queries per user). We omitted from this study all users such that the amount of recored sessions associated with them was less than $100$ (the reported numbers are calculated without those users). 
An underlying assumption is that users asking the queries are referring to themselves. Previous work suggests that this is the case in the vast majority of medical symptom search queries \cite{yomtov2015automatic}.

\subsection{Mouse Tracking}
\label{sec:mouse_tracking}
Mouse tracking data consist of a list of time-stamped horizontal ($x$) and vertical ($y$) coordinates of the cursor location during the interactions of the user, following a search query. We represent these data by extracting summarizing features.
Next, we detail the extracted features. As we explain, some of the features were manually constructed, while others were learned using representation learning. 

\paragraph*{Manually extracted features}
We represent the mouse tracking data by extracting summarizing features, such as the velocity or jerk of the pointer movement, as listed in Table \ref{tab:features}.
Another example of a summarizing feature is the number of up-down shakes. This feature provides information on how many times the user's hand was recorded shaking. 
Other features such as the minimal/maximal points on the screen were considered, but were found not to add information and were thus excluded from our analysis. 

\begin{table*}[t]
  \centering
  \begin{tabular}{|M{8.5cm}|M{7cm}|}
    \hline
    \textbf{Feature}& \textbf{Explanation}  \\
    \hline   
     Query is auto suggested&	A Boolean feature\\
     \hline 
Spell suggestion correction was done&	A Boolean feature\\
\hline 
Scroll distance&	Estimated using the recorded coordinates\\ 
\hline 
Time until the first interaction/keyboard/pointer/scroll event&	In milliseconds\\
\hline 
Total time of the session&	In milliseconds\\
\hline 
Mean/Max velocity/acceleration/jerk of $x$, $y$&	Computed using the time stamps\\
\hline 
    Number of clicked results& \\
    \hline
Number of right-left/up-down shakes&	Computed using the x coordinates\\ \hline
Longest/Average dwell time &	Computed using the time stamps \\ \hline
Number of words in the query &	\\ \hline
Number of spelling errors	& \\ \hline
Time of day&	The hour which the query was asked \\ \hline 

  \end{tabular}
  \caption{Features from users' interactions with the SERP through both mouse movements and clicks. $x$ and $y$ are the horizontal and vertical screen coordinates of the cursor, respectively.}
  \label{tab:features}
  \vspace{-24px}
\end{table*}

\paragraph*{Learning representation of mouse tracking}
As mentioned, the authors of \cite{radford2017learning} used a character-lever prediction to extract high-level features from text documents. Based on their approach, we predict event-level mouse movements in order to extract features from the raw mouse tracking data. Our feature extraction problem is, however, inherently different from the following reasons. First, time is not a factor when analyzing text document. However, in our setting, the mouse events are sampled in a non-uniform manner, and, therefore, the time gap between consecutive mouse events need to be taken into account as well (as opposed to text data where the next token is exactly one token apart from the previous one). Second, sampling mouse events is noisy, which results in missing and corrupted data.

To deal with the first problem, we predict not only the next mouse position, but also the next event time stamp, allowing our algorithm to derive the mouse velocity, acceleration and higher level derivatives as intermediate quantities. To cope with the problem of noisy and corrupted data, we use robust scaling, i.e. scaling the median of the mouse events to zero and the average absolute value of the $25$-th percentile and the $75$-th percentile to $1$.

These scaled features are then inserted into an LSTM model, followed by a fully-connected layer, which attempts to predict the raw features of the next event. The outputs of the LSTM layer at the end of the session are chosen as the representative features of this session. Importantly, note that this model is unaware of the user underlying medical state, i.e., whether she have been diagnosed with PD. Our code is publicly available\footnote{Note however that while our code is available, our data is not, for reasons of privacy.} on \cite{Code}.

\subsection{Prediction models}
Using the extracted features, we trained a classifier model that takes as an input a feature vector, corresponding to a user session, and outputs the likelihood that the user has PD. The underlying model used was a \textit{Boosted Decision Forest}, where the learning rate is set to $0.08$ and the number of trees is $100$.
To provide a single score per user based on multiple sessions, we employ majority vote as an aggregation function to decide whether a user has the condition, based on the fraction of sessions with a score greater than $0.5$.

\section{Results}
We next report the accuracy results of our model and its restricted variants, allow to asses the contribution of the learned and manually designed features. Furthermore, we analyze the accuracy of the models as a function of the amount of data collected per user. Last, we report the performance of the model predicting the next mouse location and different parameters affecting its achievements.

\subsection{Predicting PD from mouse movements}
\begin{figure}[t]
  \centering 
    \includegraphics[width=0.5\textwidth]{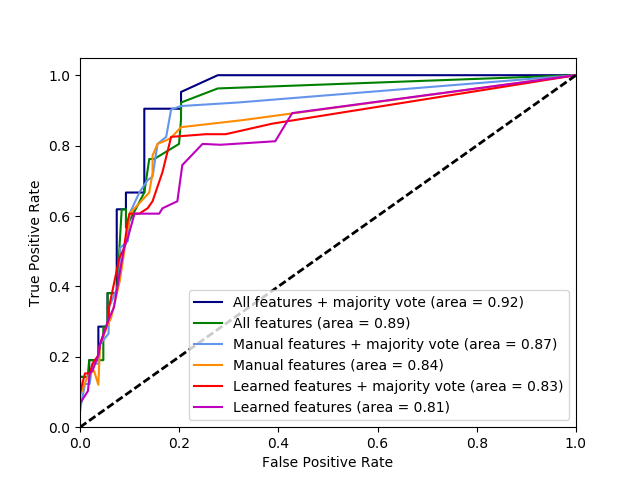}
    \vspace{-22px}
  \caption{The ROC curve and AUC of the employed models.}
    \label{fig:roc}  
    \vspace{-8px}
\end{figure}

Classifiers accuracy was assessed using the Receiver Operating Curve (ROC) and the area under it (denoted by AUC). 
We distinguish between two parameters defining the model:
(i) The features used for predictions (manually extracted, learned or both), and (ii) The performance of a model from a single session and after aggregating all sessions of a user (majority vote).
The AUC and ROC curves the classifiers are depicted in Figure \ref{fig:roc}. In all models, we used $5$-fold cross validation, while ensuring that all vectors associated with one user are within a single fold. 

First, we note that the model that uses all features and employs aggregation over users' scores is successfully able to detect PD from the users interactions (an AUC of $0.92$). The difference between the areas under the independent ROC curves was found to be not statistical significant when comparing the models that use all features with and without aggregation, yet was found to be statistically superior to all other variants (p-value $< 0.01$ for all pairs, using the Hanley \& McNeil test).  
This positive results indicate that such a technique is indeed adequate for the identification of motor disorders, and in particular, implies the possibility of using such an approach for prodromal PD identification.  

Second, observe that the performance of the models followed by the majority vote aggregation were consistently better than the ones predicting per session (although, in all settings, the difference is not statistically significant). This anticipate result indicates that such an aggregation is indeed necessary in order to make an informed decision on a user, based on sufficient data collected over time.

Interestingly, we observed that the relative contribution of the learned features compared to the manually extracted ones was indeed significant (the model that uses all features achieved an AUC of $0.92$, compared to the one that uses only expert-generated features that achieved an AUC of $0.87$). However, when measured separately, the manually constructed-features-based models outperformed the auto-learned ones, indicating the importance of using summarizing expert-generated features. Further to that, the most significant features in the model which uses all features were the time to the first interaction event, the average dwell time, and one discovered unit performing PD prediction, i.e., predicts whether the person performing the session has PD. The latter matches the success of \cite{radford2017learning}, who discovered a single unit that performs sentiment detection. 

\vspace{-2px}
\subsection{Performance as a function of data availability}
Here we focus on the model achieving the best prediction results: the one that uses all features, followed by an aggregation-per-user. We aim to quantify the effect of data availability on classification accuracy. 
Naturally, the more data we collect on users the better performance we expect, however, at a certain point, the model should saturate, with additional data having only marginal improvement. It should be pointed out that the data used is the users' search queries preformed on the Bing search engine and their corresponding collected data, and therefore, there is should be no significant difference between the amount of data collected for sick and healthy users.

\begin{figure}[!ht]
  \centering 
    \includegraphics[width=0.45\textwidth, height = 38mm]{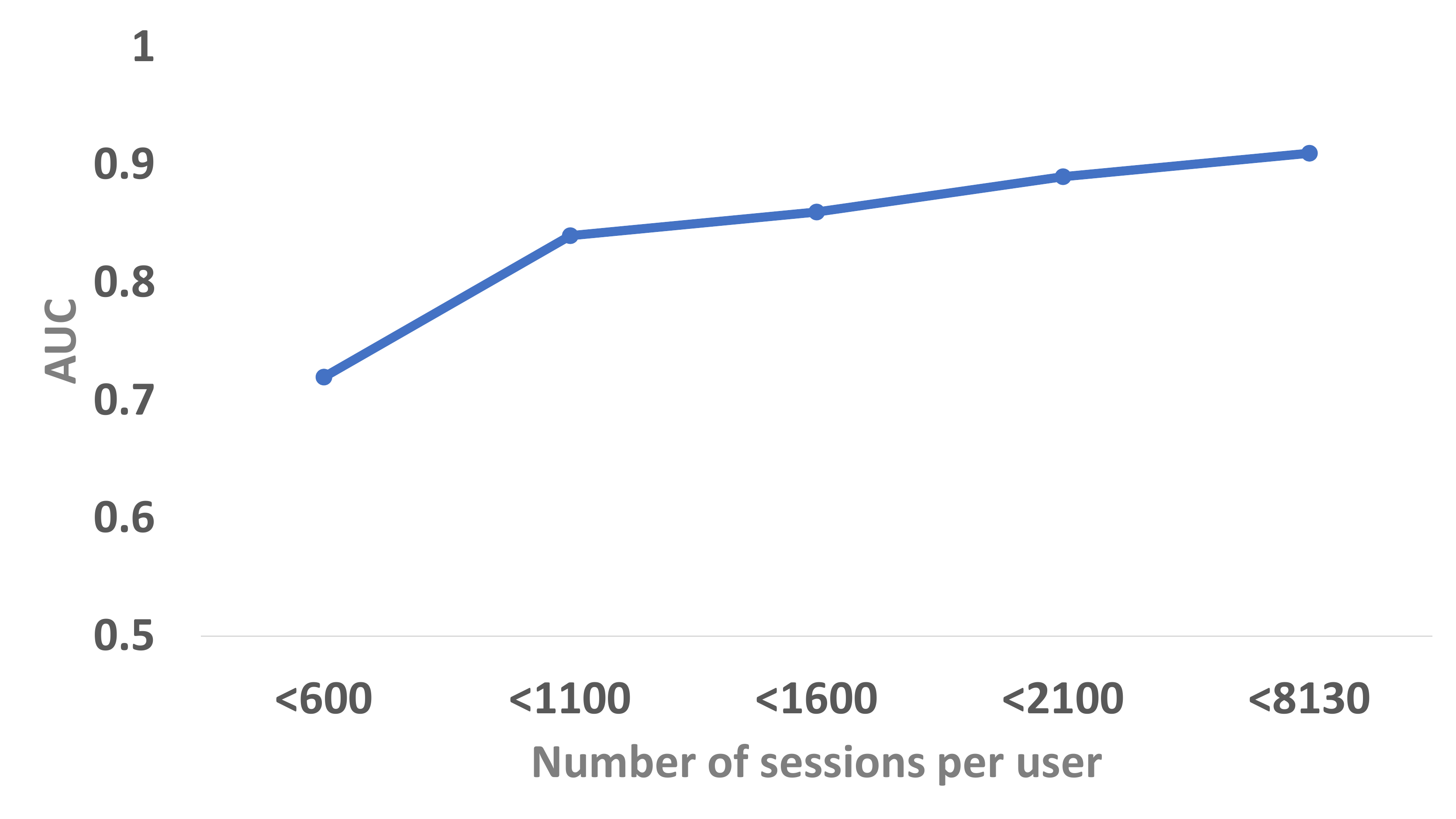}
    \vspace{-15px}
  \caption{AUC as a function of number of sessions per user.}
    \label{fig:num_sessions}  
    \vspace{-15px}
\end{figure}

For this experiment we divided the users into $7$ subsets according to their number of associated sessions, as depicted in Table \ref{tab:partition} (this partition was done in accordance with the relative amount of data per user, in leaps of $\sim$15\%). We then employed the model, first on only the first subsets, then on the first two subsets and so on. The results are presented in Figure \ref{fig:num_sessions}. 

As expected, the AUC of the model increases when more data is added (more users are also taken into account for testing, at each point on the graph). However, as one can see, at the last stage, where the largest amount of data-per-user is added, the marginal difference is small, indicating that an informed decision per user can be made even when the amount of data is limited. This result also indicates that around 1500 queries (corresponding to an average of $15$ months of queries) per person are sufficient data for tracking motor changes in users behavior and that our model is applicable even in cases where the amount of data is limited.

\begin{table}[t]
  \centering
  \begin{tabular}{|M{2cm}|M{2cm}|M{2cm}|}
    \hline
    \textbf{Range of sessions}& \textbf{Number of sick users} & \textbf{Number of healthy users}  \\
    \hline  
    100-600& 31&25\\
      \hline 
        600-1100& 88&73\\
      \hline 
      1100-1600&62&21\\
      \hline
        1600-2100&54&29\\
      \hline
        2100-8030&56&15\\
      \hline

  \end{tabular}
  \caption{Users Partition according to their sessions amount.}
  \label{tab:partition}
  \vspace{-24px}
\end{table}

\subsection{Predicting the next mouse location}
We next report the performance of the LTSM model that extracts features from mouse tracking data. 
Recall that this model, given a session's prefix, predicts the next cursor location and time stamp. Our hypothesis is that this model performs better on the ``more predictable'' healthy users than on the sick ones, as the former's mouse movements have less irregularities. 
To confirm this hypothesis, we considered the model predictions when counting a prediction within a radius of 2 pixels from the true location as a ``hit", and otherwise as a ``miss" (we ignored the time stamp predictions, as the vast majority of those predictions were very close to the real time stamps). 
Indeed, as expected, the model better predicts the healthy users' next locations than the sick users ones (AUC of $0.83$ and $0.79$).

Last, we note that in our experiments, as well as in the experiments conducted in \cite{radford2017learning}, an underlying assumption is that the data consists of samples belonging to two classes (sick/healthy, positive/negative), and is (nearly) balanced among those classes (even though the models are unaware of the labels). We examined the affect of using unbalanced training data, e.g., training the model on samples mostly belong to healthy users. Not surprisingly, we found that the data distribution is a key aspect of this approach. In particular, we found that if there are not enough samples from each class, the model is unable to learn the label as one of the features. For example, when training the model on samples mostly belonging to healthy users (90\% of the samples), the AUC of the model predicting PD which uses only the learned features dropped from $0.81$ to $0.56$.

\bibliographystyle{ACM-Reference-Format}
\bibliography{sigproc} 

\end{document}